# MUNU final results

Zornitza Daraktchieva [1]

Institute de Physique, A.-L. Breguet 1, CH-2000 Neuchâtel, Switzerland

E-mail : Zornitza.Daraktchieva@unine.ch

**Abstract**.

The MUNU detector has been designed to study $\bar{\nu}_e e^-$ elastic scattering at low energy. The central tracking detector is a 1 m$^3$ Time Projection Chamber surrounded by an anti-Compton detector. In this paper the results from final analysis of the data recorded at 3-bar and 1-bar pressure are presented. At 3-bar a new upper limit on the neutrino magnetic moment *μ$_ν$ < 9 × 10$^{-11}$(7 × 10$^{-11}$)* $\mu_B$ at 90 (68%) C.L. is derived. At 1-bar pressure electron tracks down to 80 keV are measured in the TPC. A $^{137}$Cs photopeak is reconstructed by measuring both the energy and direction of the Compton electrons in the TPC.

## 1. MUNU experiment

Technical details of the MUNU detector have already been presented in references [1, 2]. Here are given only the essentials. Briefly the detector is located at 18 m from the core of a 2.75 GW reactor in Bugey serving as an antineutrino source. The lab has overburden of steel, concrete and water corresponding to 20 meter of water. The central part of the detector is a cylindrical Time Projection Chamber (TPC). As shown in figure 1 the acrylic vessel TPC,

---

[1] On behalf of the MUNU collaboration (LPSC Grenoble, INFN Padova, Institute de Physique Neuchâtel and Physik Institut Zurich)

90 cm diameter and 160 cm long contains 11.4 (3.8 kg) of CF4 gas at 3 (1) bar pressure. The gas was chosen because of its high density, low Z, which leads to less multiple scattering and its absence of free protons, which excludes the background from the inverse beta decay. The drift volume of the TPC acts at the same time as a target for the neutrinos and as a detector for the recoiling electrons. An anode plane with 20 μm wires and pitch of 4.95 mm is used to amplify the ionization charge. The integrated anode signal gives the total deposited energy. An $x$ - $y$ plane behind the anode plane provides the spatial information of the tracks in the $x$ and $y$ directions. The third projection $z$ is obtained from the time evolution of the signal. The acrylic vessel is immersed in a steel tank (2 m diameter and 3.8 m long), filled with 10 m$^3$ of liquid scintillator (NE235) and viewed by 48 photomultipliers (PMT). The liquid scintillator acts as an anti-Compton detector and as a veto against the cosmic muons with an efficiency of 98 %. In addition, the detector is shielded against local activities by 8 cm of boron loaded polyethylene and 15 cm of Pb.

In [3] we presented an analysis of the data taken at a pressure of 3-bar, amounting to 66.6 days reactor-on and 16.7 days reactor-off. Here is presented the final analysis of 3-bar data, which takes better advantage of the electron kinematics. Also an analysis of 5.3 days of reactor-on data taken at 1-bar pressure is given.

We define a neutrino candidate as a single electron fully contained in a 42 cm fiducial radius with no energy deposited above 90 keV in the anti-Compton detector. The initial direction of the electrons is obtained by a visual fit [3]. From this fit the angles $\theta_{det}$, $\theta_{reac}$ and $\varphi_{det}$ are determined in the $x$ - $z$ and $y$ - $z$ projections. Here $\theta_{det}$ is the angle with respect to the detector axis, $\theta_{reac}$ is the scattering angle with respect to the reactor-detector axis and $\varphi_{det}$ is the angle between the projection of the initial track direction on the $x$-$y$ plane and the vertical $y$ axis(see figure 2). The background from the activities on the readout plane side of the TPC is suppressed by applying the angular cut *$\theta_{det}$ < 90°*[3].

## 2. Results from 3-bar forward-normalized background analysis

At 3-bar we select electron events with kinetic energies $T_e > 700$ keV. For each electron track the neutrino energy $E_\nu$ is reconstructed from the scattering angle $\theta_{reac}$ and measured electron recoil energy $T_e$. The selected events with $E_\nu > 0$ and beginning of the track within a forward cone are the forward electrons (figure 2). The axis of the forward cone coincides with the reactor-detector axis. In the same way the electrons in the tree background cones: backward, upward and downward are selected to estimate the background. The backward cone is defined as opposite to the forward cone while upward and downward cones are perpendicular. To avoid overlap of the cones, which can occur for $T_e < 2m_ec^2$, we require in addition that the angle $\varphi_{det}$ is within less then ± 45° of the cone axis.

The forward electrons are signal plus background events. The backward, upward and downward electrons are all background events. The background rates in the backward, upward and downward cones are normalised by dividing by 3 to the rate in the forward cone. This normalized background (NB) is then directly compared with the event rate in the forward cone [4].

The energy distributions of both forward (455 ± 21) and NB (384 ± 11) electrons are given in figure 3. There is a clear excess of the events in forward direction from $\bar{\nu}_e e^-$ scattering. The total forward minus NB above 700 keV is 71 ± 23 counts for 66.6 days live time reactor-on. We make the same analysis with the data taken during the reactor-off period as a cross check (16.7 days live time). The energy distributions of both forward (133 ± 11) and NB electrons (147 ± 7) are given in figure 4. The integrated forward minus NB rate above 700 keV is -0.8 ± 0.8 counts per day (cpd), consistent with zero.

Now we estimate the expected rate from weak interactions. The Monte Carlo simulations (GEANT 3) are used for calculation of various acceptances of the electron selection procedure [3]. The detector containment efficiency in the 42 cm fiducial radius was found to vary from 63 % at 700 keV, 50 % at 1 MeV to 12 % at 2 MeV, with an error of 2 %.

The reactor neutrino spectrum was calculated using the procedure described in [3, 6]. The uncertainties in the neutrino spectrum are 5 % above 900 keV and 20 % below that. The errors on the global acceptance, including track reconstruction efficiency (4 %), are of order of 7 %. The expected event rate above 700 keV is found to be 1.02 ± 0.1 cpd, in good agreement with the measured one 1.05 ±0.36 (cpd).

Both the measured and expected rates are displayed in figure 5. The large excess of events in the first two bins (700 – 900 keV) observed in our previous analysis has to a large extend disappeared. It is thus most likely due to a statistical fluctuation in the background, more precisely determined in the present analysis. $\chi^2$ method was used as in [3], with the same binning to compare the measured and expected spectra. The minimum $\chi^2$ is 11.5 for 7 degrees of freedom for a squared magnetic moment $\mu^2 = -0.72 \times 10^{-20} \mu_B^2$. We find in 90 % confidence interval $\mu^2 = (-0.72 \pm 1.25) \times 10^{-20} \mu_B^2$. This result is compatible with the weak interaction alone, and there is no indication of a magnetic moment. To estimate the limit on the magnetic moment we renormalize the result to the physical region ($\mu^2 > 0$) and obtain the limit $\mu_\nu < 9(7) \times 10^{-11} \mu_B$ at 90 (68) % C.L.

## 3. 1-bar analysis

### 3.1. Results from 1 bar forward – normal background analysis

During the 1-bar data taking period the threshold on the electron recoil was lowered to 100 keV. Events in the TPC which are not in coincidence with a gamma above 90 keV in the scintillator in an 80 μs time window are accepted. The neutrino trigger rate is 0.4 Hz with 50 % deadtime, mostly due to the data read-out and data transfer time. The measurements with radioactive sources showed that the energy resolution of the TPC at 1-bar is following an empirical $T^{0.57}$ law, being about 2 times better than the energy resolution at 3-bar [5].

An example of a low energy single electron of 170 keV in 1bar of CF4 is displayed in figure 6. For single electrons fully contained inside the TPC above 200 keV the rate is 760 cpd/kg. We used the same kinematical procedure as abovementioned to determine the background and to select the neutrino candidates. After applying the fiducial, geometrical and kinematical cuts we have measured the signal from neutrino electron interaction, corresponding to 2.89 ± 2.39 counts per day.

**3.2. Reconstruction of Cs photopeak**

Here is given the reconstruction of the incident photon energy obtained with the 662 keV $^{137}$Cs source at 1 bar of CF4. The photons interact with electrons in the TPC via Compton scattering. The scatter photon is measured in the scintillator. The recoil electron track and energy are measured in the TPC. The electron direction is coinciding with the incident photon direction in the first centimetres of the track. The Compton electrons initial direction is obtained with a visual fit. The angle $\theta_{source}$ which is the angle with respect to the source axis (being perpendicular to detector-reactor axis) is calculated from this fit. The initial photon energy $E_\gamma$ is reconstructed from the scattering angle $\theta_{source}$ and the electron recoil energy $T_e$, measured in the TPC. The reconstructed photopeak is compared with Monte Carlo simulations in figure. 7. The width of the peak at 1$\sigma$ is 220 keV at 662 keV. The angular resolution of the Compton recoil spectrum above 150 keV is $\sigma_\theta = 11.6° \pm 0.9$.

## 4. Conclusion

The MUNU experiment studied neutrino electron scattering at low energy using a nuclear reactor as an antineutrino source. Both the energy and scattering angle are measured. A reduction of the statistical errors is achieved due to a better estimation of the background in the entire chamber. A good agreement is seen between the measured and expected from weak interactions spectra above 700 keV for operation of 3-bar. A limit of $\mu_v < .9 \times 10^{-11} \mu_B$ at 90 % CL is derived. At 1 bar pressure, the direction and energy of the electron tracks are reconstructed above 200 keV. Also at 1 bar, a 662 keV $^{137}$Cs photopeak is reconstructed from Compton scattering.

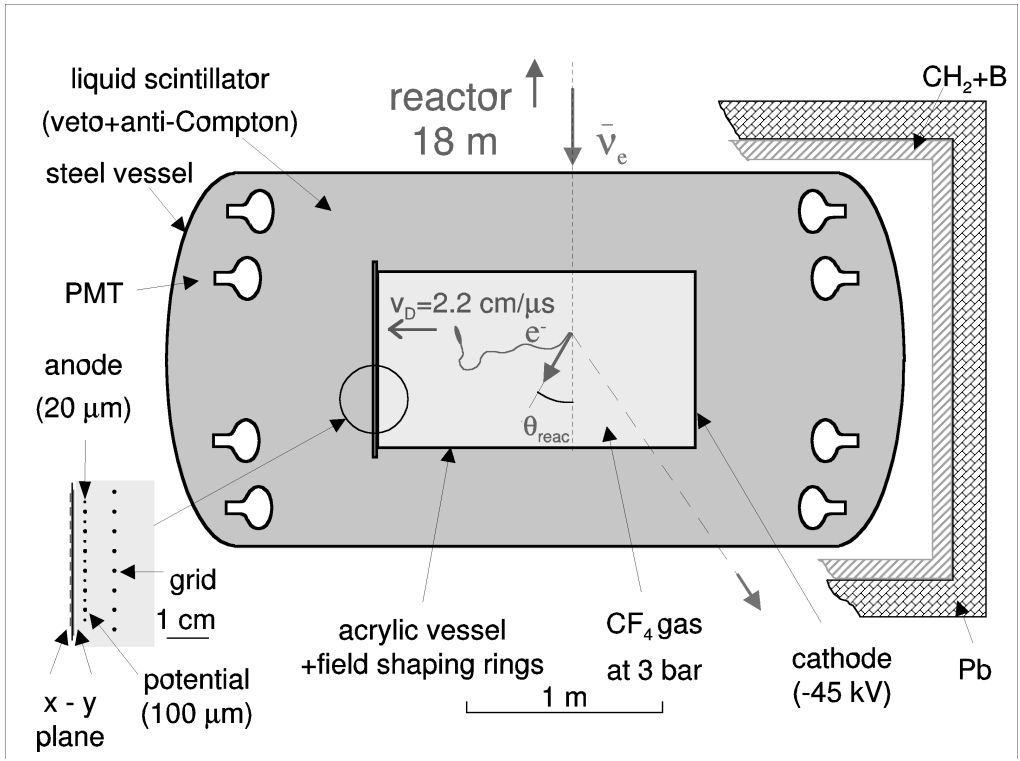

**Figure 1.**

The MUNU detector

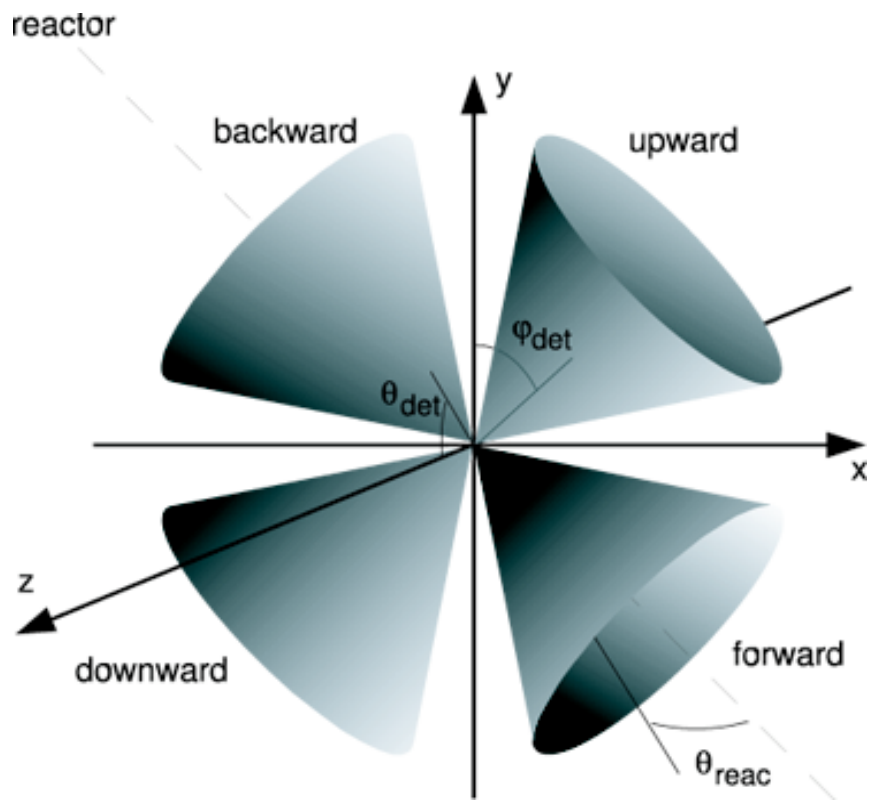

**Figure 2.**

The four kinematical cones

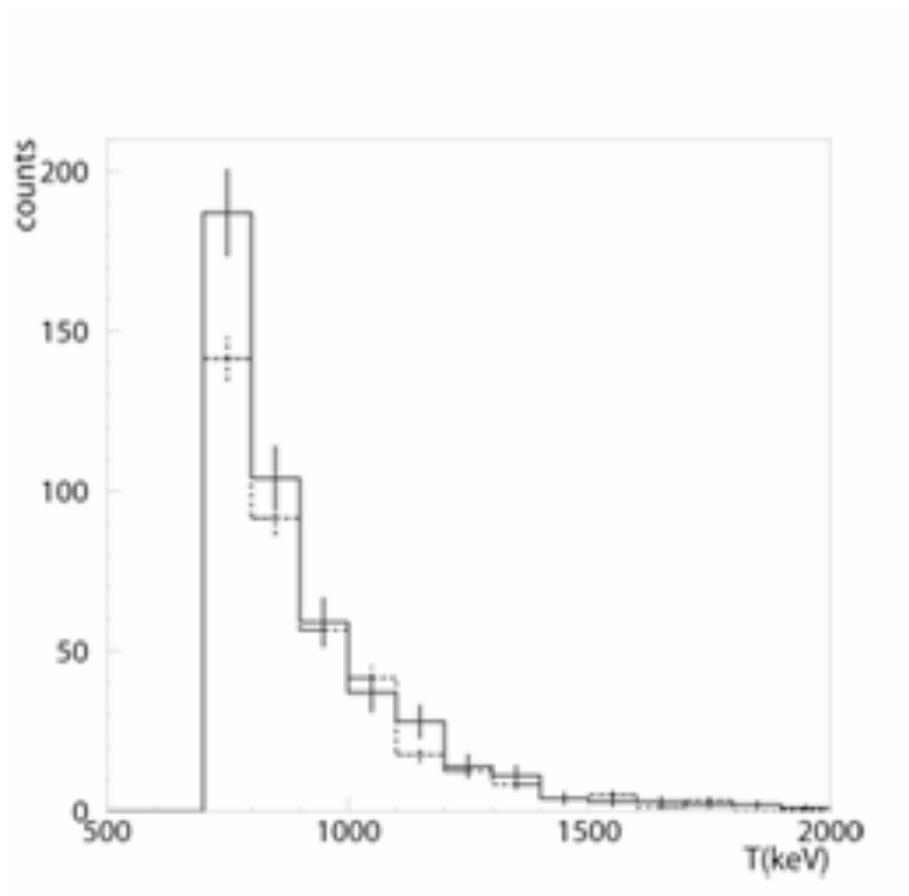

**Figure 3.**

Forward (solid line) and NB (dashed line) electrons, 3 bar, reactor on

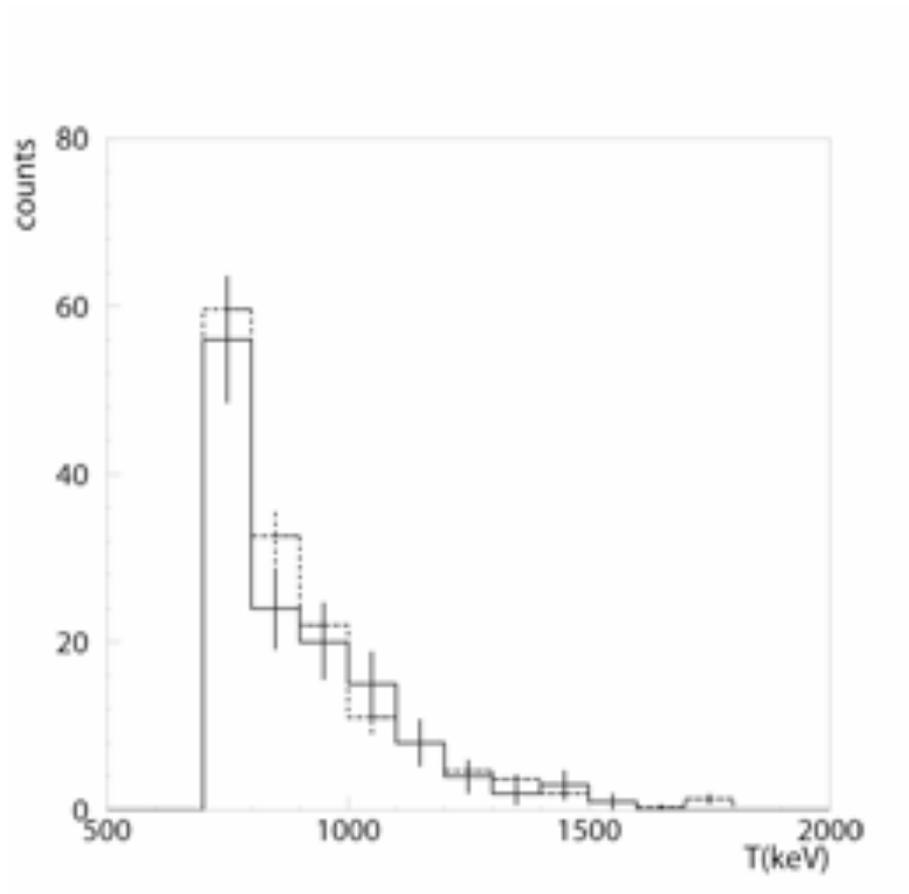

**Figure 4.**

Forward (solid line) and NB (dashed line) electrons, 3 bar, reactor off

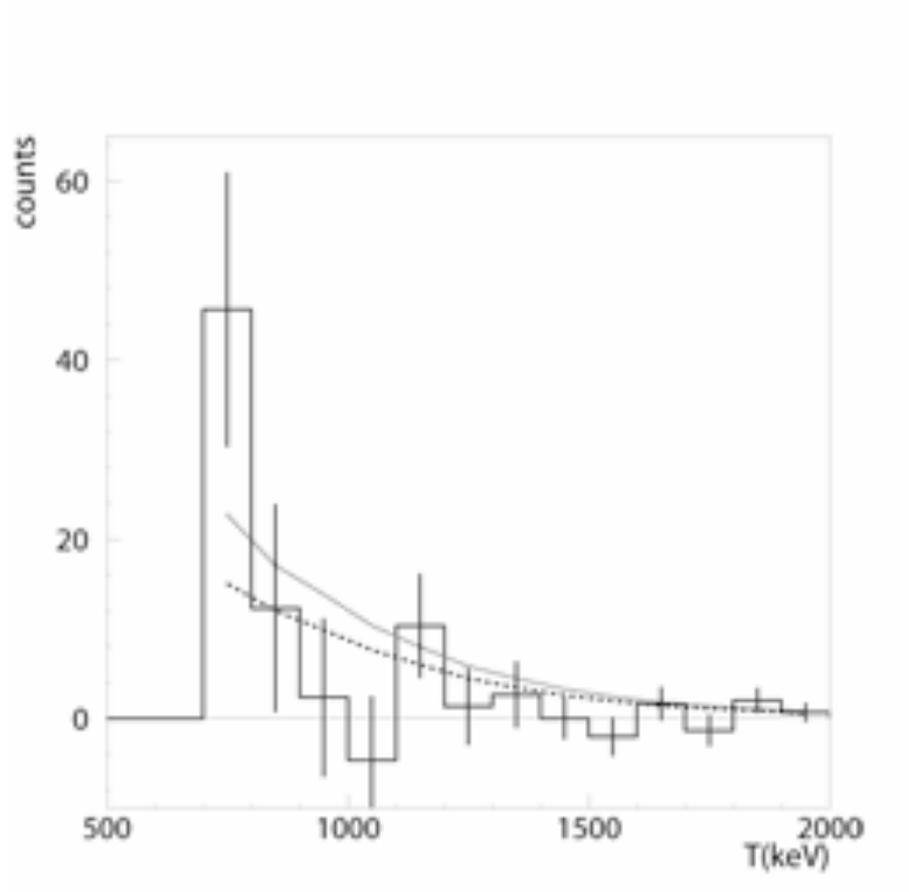

**Figure 5.**

Energy distribution of forward minus NB electrons, 3 bar, reactor on (solid line), expected spectrum from weak interactions alone (dashed line) and estimated for a magnetic moment of $9 \times 10^{-11} \mu_B$ dotted line).

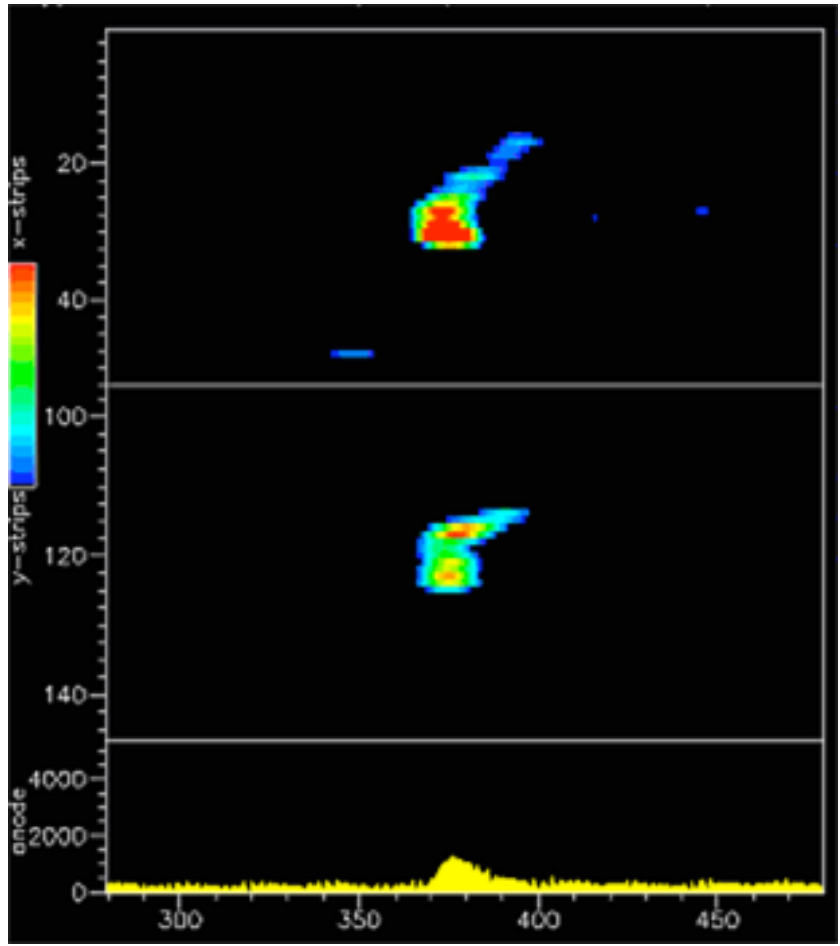

**Figure 6.**

A 170 keV electron in the TPC at 1 bar of CF4: *x – z*, *y –z* projections as well as the integrated anode signal are displayed as a function of time (*z*). The binning is 3.5 *mm* for *x* and *y* and 80 *ns* for *z*.

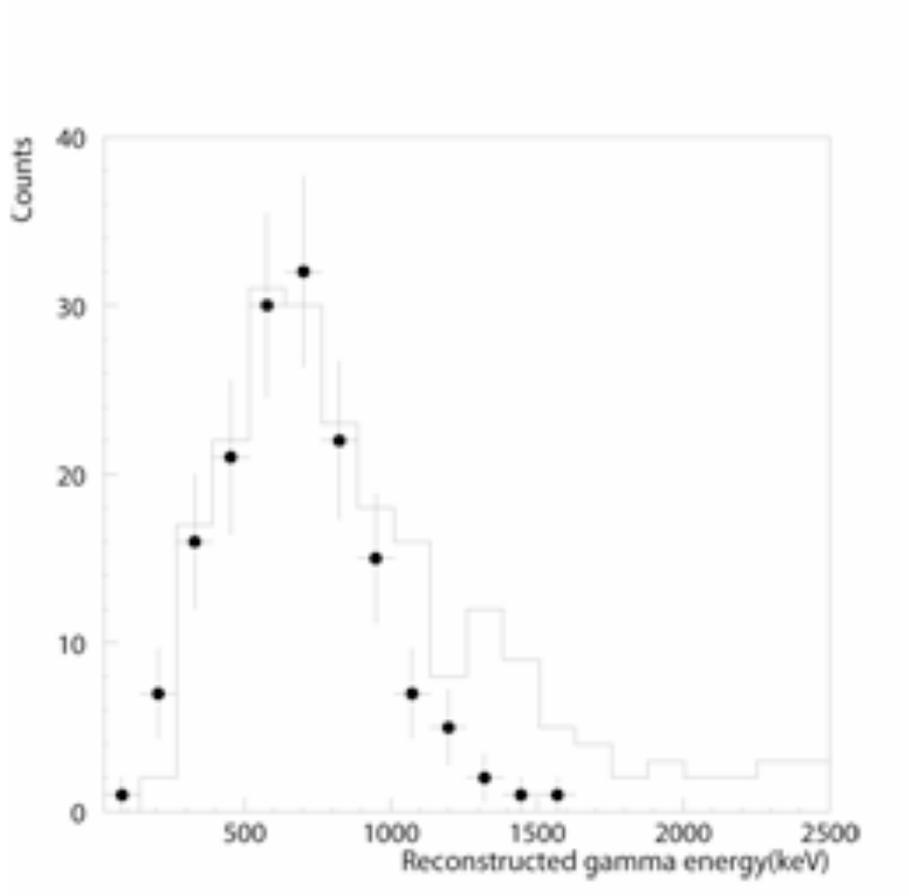

**Figure 7.**

Reconstructed $^{137}$Cs photopeack of 662 keV